# Geometric phase of a bipartite system with Dzyaloshinski-Moriya interaction

Yue Zhou[1], Guo-Feng Zhang[2]*

[1]*State Key Laboratory for Superlattices and Microstructures, Institute of Semiconductors, Chinese Academy of Sciences, P. O. Box 912, Beijing 100083, People's Republic of China*

[2]*Department of Physics, School of Sciences, Beijing University of Aeronautics & Astronautics, Xueyuan Road No. 37, Beijing 100083, People's Republic of China*

**Abstract**

The Berry phase of a bipartite system described by a Heisenberg *XXZ* model driven by a one-site magnetic field is investigated. The effect of the Dzyaloshinski-Moriya (DM) anisotropic interaction on the Berry phase is discussed. It is found that the DM interaction affects the Berry phase monotonously, and can also cause sudden change of the Berry phase for some weak magnetic field cases.

The geometric phase, which was first studied by Berry, has attracted much interest these days. When a time-dependent quantum system undergoes a cyclic and adiabatic evolution, an unintegrable phase, which is called Berry phase, is attached to the wave function. Though the Berry phase had been discovered independently for numerous times (e.g. [1-4]), its properties and physical meaning were not truly comprehended prior to Berry [5] and Simon's [6] work. The geometric phase that Berry considered originally was limited to adiabatic and cyclic evolution. Aharonov and Anandan [7] expanded it to the non-adiabatic cases. And then, Samuel and Bhandari [8] generalized it further to non-cyclic, and even non-unitary process.

As an important evolvement, the application of the geometric phase has been proposed in many fields, such as the geometric quantum computation [9]. Due to the global properties of the geometric phase, the local random fluctuation is negligible, which is propitious to construct fault-tolerant quantum gates. A number of physical systems have been investigated to realize geometric quantum computation, such as NMR (Nuclear Magnetic Resonance), Josephson junction, Ion trap and semiconductor quantum dots [10]. In the proposed schemes, the systems generally consist of two weakly coupled spins with one or both spins driven by a time-varying magnetic field. The vector locus of the driving magnetic field forms a closed path, so that the geometric phase is acquired.

Due to the ubiquity of the anisotropic interaction, it is necessary to take the influence of anisotropy into account. Sun *et al.* studied the effect of spin-spin coupling based on the Heisenberg *XXZ* model [11], and the more generalized *XYZ* type interaction was discussed by perturbation method in Ref. [12]. Although M. Asoudeh [13] stated that the different types of anisotropic interactions may not be of much practical relevance to concrete physical realization of qubits, the solid state structure is intrinsically inhomogeneous and magnetic imperfections or impurities are likely to be present, which lead to spin-orbit coupling anisotropic interactions. Recently, the physics of semiconductors with a spin-orbit interaction has attracted much attention, as it plays an important role in the emerging field of semiconductor spintronics [14]. Sun *et al*. investigated the equilibrium property of a mesoscopic ring with a spin-orbit interaction and studied the persistent spin current [15]. In view of the above results, we think it would be interesting to investigate the effects of anisotropic interactions on the Berry phase.

---

* Corresponding author.
  *Tel:* +86-010-82317935
  *E-mail address:* gf1978zhang@buaa.edu.cn

In this paper, we study the influence of the spin-orbit coupling on the Berry phase. The system is described by an *XXZ* model with one-site magnetic drive in the presence of the Dzyaloshinski-Moriya (DM) anisotropic antisymmetric interaction. The Hamiltonian we consider in this paper is given by

$$H = \frac{1}{2}[J(\sigma_1^x \sigma_2^x + \sigma_1^y \sigma_2^y) + J_z \sigma_1^z \sigma_2^z + \vec{B} \cdot \vec{\sigma}_1 + \vec{D} \cdot (\vec{\sigma}_1 \times \vec{\sigma}_2)], \tag{1}$$

where $J$ and $J_z$ are the real coupling coefficients, the model is called antiferromagnetic for $J > 0$ and ferromagnetic for $J < 0$. Here we consider the antiferromagnetic case only and set $J = 1$, while the ferromagnetic case can be studied similarly. $\sigma_j^i (i = x, y, z; j = 1, 2)$ are the Pauli operators of the $j^{th}$ spin. $\vec{B} = B\hat{n}(t)$ is the driving magnetic field and $\hat{n} = (\sin\theta\cos\phi, \sin\theta\sin\phi, \cos\theta)$ is the unit vector. $\vec{D} \cdot (\vec{\sigma}_1 \times \vec{\sigma}_2)$ is the DM coupling term that arises from spin-orbit coupling [16, 17]. Choosing $\vec{D} = D\vec{z}$, then Hamiltonian (1) becomes

$$H = \frac{1}{2}[J(\sigma_1^x \sigma_2^x + \sigma_1^y \sigma_2^y) + J_z \sigma_1^z \sigma_2^z + \vec{B} \cdot \vec{\sigma}_1 + D(\sigma_1^x \sigma_2^y - \sigma_1^y \sigma_2^x)], \tag{2}$$

in the standard basis $\{|11\rangle, |10\rangle, |01\rangle, |00\rangle\}$ ($|1\rangle \equiv |\uparrow\rangle$, $|0\rangle \equiv |\downarrow\rangle$), the Hamiltonian (2) can be written as

$$H = \frac{1}{2}\begin{pmatrix} J_z + B\cos\theta & 0 & B\sin\theta e^{-i\varphi} & 0 \\ 0 & -J_z + B\cos\theta & 2J + 2Di & B\sin\theta e^{-i\varphi} \\ B\sin\theta e^{i\varphi} & 2J - 2Di & -J_z - B\cos\theta & 0 \\ 0 & B\sin\theta e^{i\varphi} & 0 & J_z - B\cos\theta \end{pmatrix}. \tag{3}$$

By the secular equation, it follows that the eigenvalues $E_i (i = 1, 2, 3, 4)$ of Hamiltonian (2) satisfy

$$16E_i^4 + SE_i^2 + LE_i + C = 0, \tag{4}$$

where

$$S = -8(2J^2 + J_z^2 + 2D^2 + B^2),$$

$$L = 16J_z(J^2 + D^2),$$

$$C = (B^2 + J_z^2)^2 - 4D^2 J_z^2 - 4J^2 J_z^2 + 4B^2 \cos^2\theta(J^2 + D^2 - J_z^2). \tag{5}$$

The expression of $E_i$ can be solved analytically, however, due to the complexity, it is not necessary to demonstrate their expressions here. Without the external magnetic field, the eigenvalues can be expressed

simply as

$$E_1 = -\frac{J_z}{2} - \sqrt{J^2 + D^2}, \qquad E_2 = -\frac{J_z}{2} + \sqrt{J^2 + D^2}, \qquad E_3 = E_4 = \frac{J_z}{2}, \qquad (6)$$

in which $E_3$ and $E_4$ are degenerate and independent on $D$; and in some cases, all the three excited states $E_2$, $E_3$ and $E_4$ are threefold degenerate. The one-site driving magnetic field put on the first spin consists of three parameters, the magnitude $B$, the elevation $\theta$ and the azimuth $\varphi$, each of which is controllable. Here we keep $B$ and $\theta$ constant, and $\varphi$ changes from 0 to $2\pi$ slowly, so that the magnetic field rotates around the z axis and the cyclic and adiabatic condition is satisfied. Also, since the time-independent part of Hamiltonian (2) is $\Re_2$ invariant, the eigenvalues of Hamiltonian (2) are constant during the evolution. When the driving magnetic field is put on, all the degeneracy is removed. The stronger the magnetic field is, the wider the energy level intervals are, as shown in Fig. 1. Besides, all the four eigenvalues are now dependent on the value of $D$, which are plotted in Fig. 2.

After some straightforward algebra, we can obtain the instantaneous eigenstates of Hamiltonian (2):

$$|\psi_i\rangle = \frac{1}{U_i}(a_i|11\rangle + b_i|10\rangle + c_i|01\rangle + d_i|00\rangle, \qquad (7)$$

in which

$$a_i = B\sin\theta e^{-i\varphi},$$

$$b_i = \frac{4E_i^2 - J_z^2 - B^2 - 2J_z B\cos\theta}{2J - 2Di},$$

$$c_i = 2E_i - J_z - B\cos\theta,$$

$$d_i = \frac{B\sin\theta e^{i\varphi} b_i}{2E_i - J_z + B\cos\theta},$$

$$U_i = \sqrt{|a_i|^2 + |b_i|^2 + |c_i|^2 + |d_i|^2}. \qquad (8)$$

With the instantaneous eigenvectors, the Berry phase of the system can be routinely written as

$$\beta_i = i\int_0^t \left(\psi_i(t'), \frac{\partial \psi_i(t')}{\partial t'}\right) dt'. \qquad (9)$$

Note that of the coefficients in Eq. (8) only $a_i$ and $d_i$ are time-dependent, it follows that

$$\beta_i = i\int_0^{2\pi} \langle \psi_i | \frac{d}{d\varphi} | \psi_i \rangle d\varphi = \frac{2\pi}{U_i^2}(|a_i|^2 - |d_i|^2). \qquad (10)$$

The Berry phase as a function of the DM interaction $D$ and the elevation $\theta$ is plotted in Fig. 3. It can be seen that the Berry phase is vanishing for $\theta = \frac{\pi}{2}$, and the image is odd symmetrical along the axis that $\theta = \frac{\pi}{2}$, $\gamma = 0$. As $D$ increases, the Berry phase increases or decreases monotonously and gently.

The Berry phase as a function of the DM interaction $D$ and the magnetic field $B$ is plotted in Fig. 4. A notable characteristic of the image is that when the magnetic field is weak there is a steep step in the $\gamma - D$ curve. Viewed physically, the sudden change of the Berry phase corresponds to crossover of the energy levels. In the system we discuss, accidental degeneracy does not exist with a nonvanishing magnetic field; however, if the magnetic field is weak, sub-degeneracy may occur in some cases, as shown in Fig. 5. Since when $B$ is small, the magnetic field term can be treated as perturbation, so that the sub-degenerate point corresponds to the threefold degenerate point in Eq. (6), where $E_2 = E_3 = E_4$. It follows that the sub-degenerate point or the sudden change point of the Berry phase satisfies the relation

$$D = \sqrt{J_z^2 - J^2} \ . \tag{11}$$

Eq. (11) shows that if $J_z < J$, the sub-degeneracy is inexistent and there is no sudden change of the Berry phase. Also, due to the fact that $D$ is generally small compared with the coupling constant $J$, in the strong anisotropy cases where $J_z \gg J$, Eq. (11) can not be satisfied either. The effect of DM interaction $D$ is most remarkable in the cases that the value of $J_z$ is slightly larger than that of $J$.

In conclusion, we have investigated the Berry phase of the bipartite system described by a Heisenberg *XXZ* model driven by a one-site magnetic field. The effect of the DM anisotropic antisymmetric interaction is analyzed quantitatively. It is found that the Berry phase changes monotonously as the DM interaction $D$ increases, and in some weak magnetic field cases, the DM interaction can cause sudden change of the Berry phase. The results may be useful in the field of geometric quantum computation.

**Acknowledgements**

This work is supported by the National Science Foundation of China under Grant No. 10604053 and the Beihang Lantian Project.

**Figure captions**

1. $B$-dependence of instantaneous eigenvalues with $J = J_z = 1$, $\theta = \pi/4$ and $D = 0.5$.

2. $D$-dependence of instantaneous eigenvalues with $J = J_z = 1$, $\theta = \pi/4$ and $B = 1$.

3. (Color online) Berry phase as a function of the DM interaction $D$ and the elevation $\theta$ when $J = 1$, $J_z = 1.1$ and $B = 1$. The Berry phase and the elevation $\theta$ are plotted in units of $\pi$.

4. (Color online) Berry phase as a function of the DM interaction $D$ and the magnetic field $B$ when $J = 1$, $J_z = 1.1$ and $\theta = \pi/4$. The Berry phase are plotted in units of $\pi$.

5. $D$-dependence of instantaneous eigenvalues with $J = 1$, $J_z = 1.1$, $\theta = \pi/4$ and $B = 0.02$. $E_1$ is not plotted here.

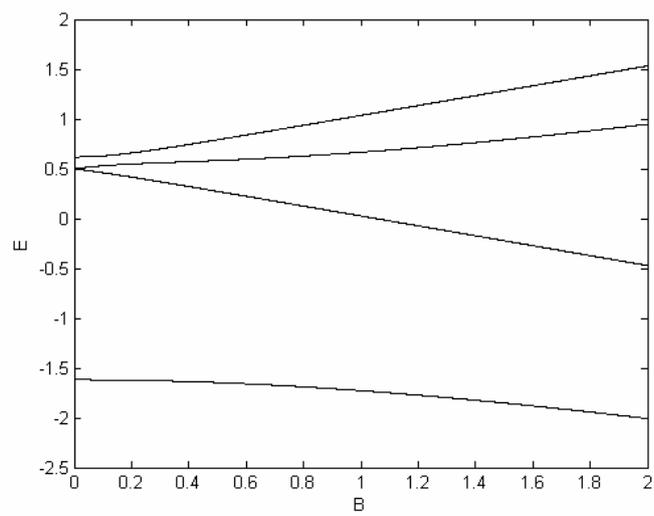

Fig. 1. $B$-dependence of instantaneous eigenvalues with $J = J_z = 1$, $\theta = \pi/4$ and $D = 0.5$.

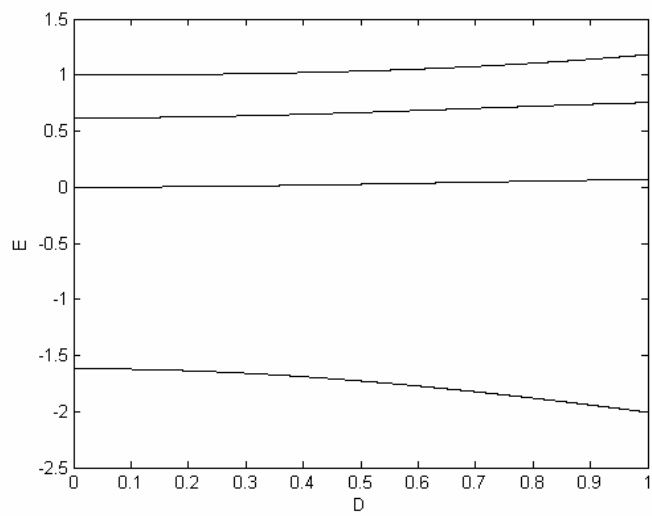

Fig. 2. $D$-dependence of instantaneous eigenvalues with $J = J_z = 1$, $\theta = \pi/4$ and $B = 1$.

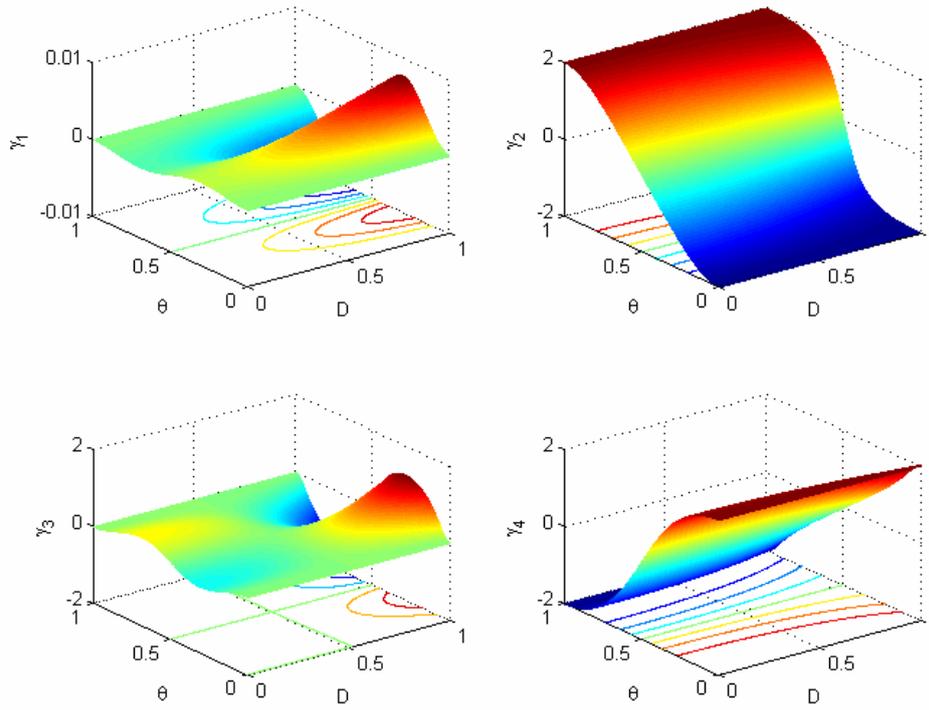

Fig. 3. (Color online) Berry phase as a function of the DM interaction $D$ and the elevation $\theta$ when $J=1$, $J_z=1.1$ and $B=1$. The Berry phase and the elevation $\theta$ are plotted in units of $\pi$.

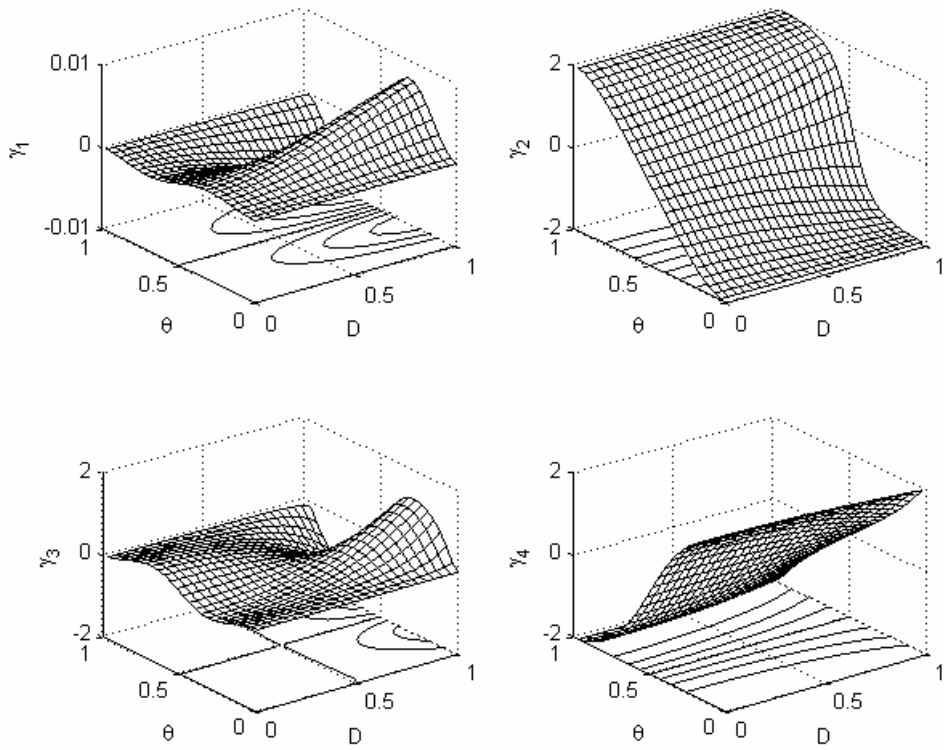

Fig. 3. Berry phase as a function of the DM interaction $D$ and the elevation $\theta$ when $J=1$, $J_z=1.1$ and $B=1$. The Berry phase and the elevation $\theta$ are plotted in units of $\pi$.

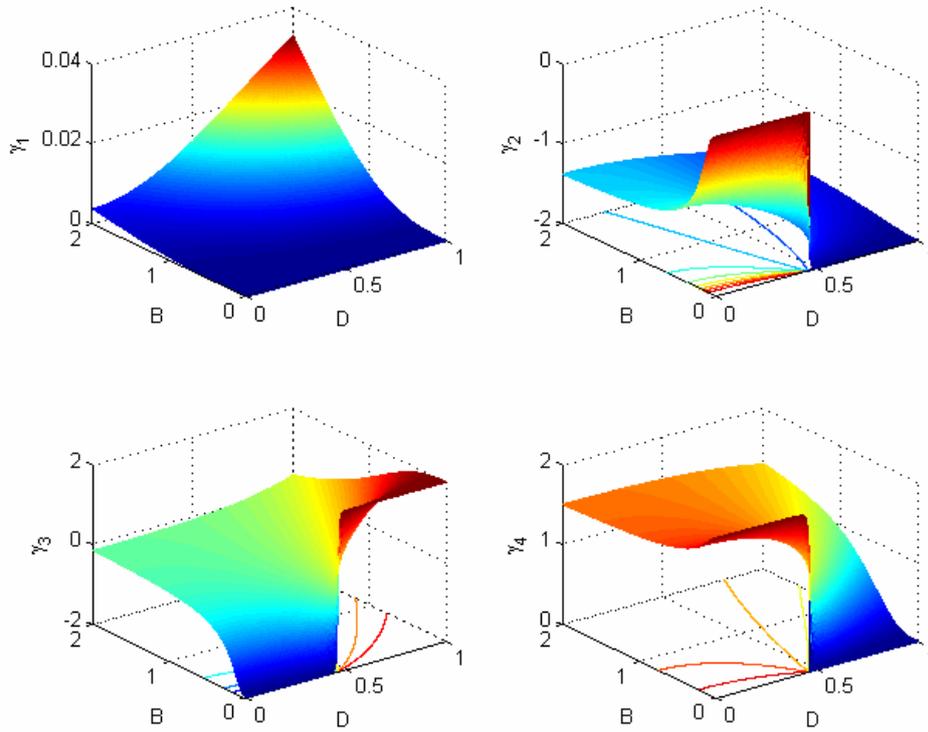

Fig. 4. (Color online) Berry phase as a function of the DM interaction $D$ and the magnetic field $B$ when $J = 1$, $J_z = 1.1$ and $\theta = \pi/4$. The Berry phase are plotted in units of $\pi$.

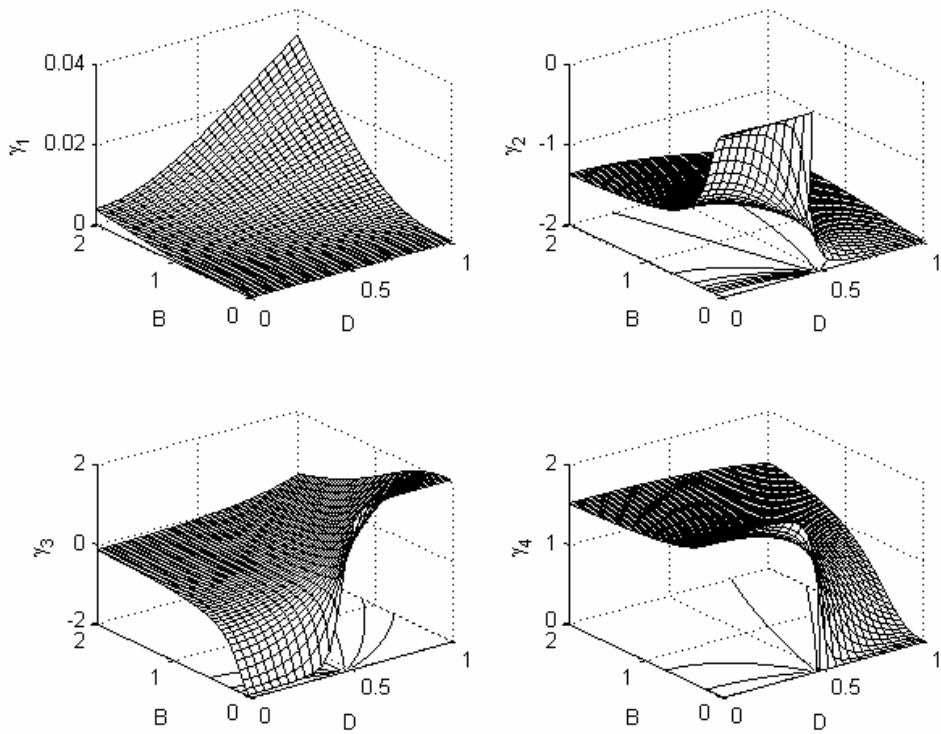

Fig. 4. Berry phase as a function of the DM interaction $D$ and the magnetic field $B$ when $J=1$, $J_z=1.1$ and $\theta=\pi/4$. The Berry phase are plotted in units of $\pi$.

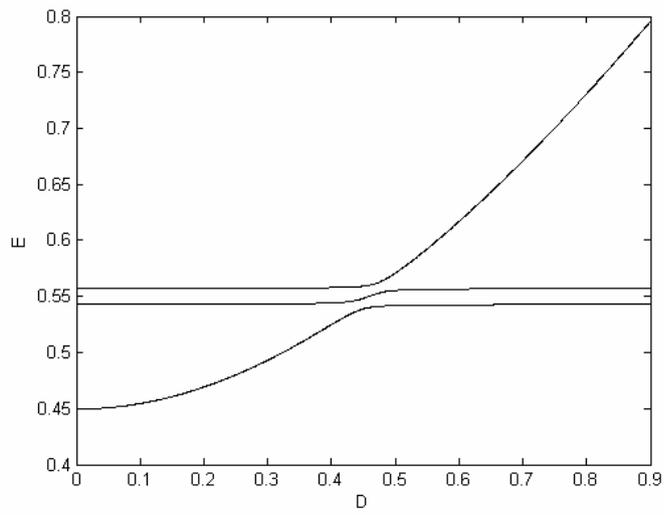

Fig. 5. $D$-dependence of instantaneous eigenvalues with $J=1$, $J_z=1.1$, $\theta=\pi/4$ and $B=0.02$. $E_1$ is not plotted here.